# Supersolid symmetry breaking from compressional oscillations in a dipolar quantum gas

L. Tanzi*[1,3], S. M. Roccuzzo*[2], E. Lucioni[1,3], F. Famà[1], A. Fioretti[1], C. Gabbanini[1], G. Modugno[1,3], A. Recati[2], and S. Stringari[2]
*These authors contributed equally to this work.

[1]CNR-INO, Sede Secondaria di Pisa, 56124 Pisa, Italy
[2]CNR-INO BEC Center and Dipartimento di Fisica, Università di Trento, 28123 Povo, Italy
[3]LENS and Dipartimento di Fisica e Astronomia, Università di Firenze, 50019 Sesto Fiorentino, Italy

**The existence of a paradoxical supersolid phase of matter, possessing the apparently incompatible properties of crystalline order and superfluidity, was predicted 50 years ago[1-3]. Solid helium was the natural candidate, but there supersolidity has not been observed yet, despite numerous attempts[4-7]. Ultracold quantum gases have recently shown the appearance of the periodic order typical of a crystal, due to various types of controllable interactions[8-12]. A crucial feature of a D-dimensional supersolid is the occurrence of up to D+1 gapless excitations reflecting the Goldstone modes associated with the spontaneous breaking of two continuous symmetries: the breaking of phase invariance, corresponding to the locking of the phase of the atomic wave functions at the origin of superfluid phenomena, and the breaking of translational invariance due to the lattice structure of the system. The occurrence of such modes has been the object of intense theoretical investigations[1,13-17], but their experimental observation is still missing. Here we demonstrate the supersolid symmetry breaking through the appearance of two distinct compressional oscillation modes in a harmonically trapped dipolar Bose-Einstein condensate, reflecting the gapless Goldstone excitations of the homogeneous system. We observe that the two modes have different natures, with the higher frequency mode associated with an oscillation of the periodicity of the emergent lattice and the lower one characterizing the superfluid oscillations. Our work paves the way to explore the two quantum phase transitions between the superfluid, supersolid and solid-like configurations that can be accessed by tuning a single interaction parameter.**

Given the lack of observation of supersolidity in solid helium[6], quantum gases with spin-orbit coupling[8,18], cavity-mediated interactions[9,19] and long-range dipolar interactions[16,20-22,10-12,17] are emerging as interesting alternatives. A major difference with respect to solid helium is that the lattice structure of these novel configurations is built by clusters rather than by single atoms, thereby naturally allowing for the emergence of coherence effects typical of a superfluid. The recently observed supersolid stripe phase in elongated dipolar Bose-Einstein condensates[10-12] is particularly appealing, since the lattice-like structure is determined directly by the atom-atom interactions and not by light-mediated interactions, allowing the lattice to be deformable. However, how to reveal the spontaneous symmetry breaking in such trapped, inhomogeneous type of systems, and how to study the lattice dynamics are still open questions.

In this work we demonstrate, both theoretically and experimentally, that the peculiar symmetry breaking of the dipolar supersolid can be revealed by studying the collective oscillations of the system in the trap. Low frequency compressional modes emerge naturally from the hydrodynamic equations of superfluids and have been studied extensively in the context of trapped Bose-Einstein condensates[23-25]. The hydrodynamic equations are the direct consequence of the locking of the phase of the order parameter and are hence peculiar of superfluid systems at low temperature[26]. We now discover that, when the system is driven from the usual superfluid to the supersolid regime by tuning the interactions, the lowest compressional mode, the so-called axial breathing mode, bifurcates into two distinct excitations, similarly to the bifurcation of the gapless Goldstone excitations expected for a homogeneous supersolid. By further varying the interactions, one of the two mode disappears, and the system shows a second transition from the supersolid to a droplet crystal regime.

The system we study is initially an ordinary Bose-Einstein condensate (BEC) of strongly magnetic Dy atoms, see Fig.1a, held in a harmonic potential with frequencies $\omega_x$, $\omega_y$ and $\omega_z$, with the dipoles aligned along $z$ by a magnetic field $B$; the potential is elongated in the $x$ direction. At a mean field level, the atoms interact with a short range repulsion parametrized by the s-wave scattering length $a_s$, and with a long range dipolar interaction parametrized by the dipolar length $a_{dd}$, which is repulsive along the $x$ and $y$ directions and attractive along the $z$ direction. Changing $a_s$ via a Feshbach resonance we control the relative interaction strength, $\epsilon_{dd} = a_{dd}/a_s$. By increasing $\epsilon_{dd}$ from the BEC side, a roton minimum develops in the excitation spectrum[27]. When the roton energy gap vanishes, for $\epsilon_{dd} = 1.38$, an instability develops and a periodic density modulation appears along the $x$ direction, with a characteristic momentum set by the trap confinement along $z$, $k_r \sim (\hbar/m\,\omega_z)^{-1/2}$ [28]. The modulation would not be stable at the mean-field level but is stabilized by the Lee-Huang-Yang interaction (LHY), which is the leading term accounting for beyond mean field quantum fluctuations[29]. For a narrow range of values of $\epsilon_{dd}$, the system shows both density modulation and phase coherence, which are the prerequisites for supersolidity[10-12]. Due to the presence of three-body recombination, enhanced by the emergence of high-density regions, the lifetime of the supersolid is limited to hundreds of milliseconds. For larger values of $\epsilon_{dd}$, the system enters an incoherent droplet crystal regime[30,31].

Theoretically, we study the system by means of a density functional theory with a local density approximation for the equation of state, which includes the LHY correction. We consider the ideal scenario of zero temperature and absence of three-body recombination. We prepare the system in its ground state at a specific value of $\epsilon_{dd}$. We induce an axial breathing mode along the $x$ direction, through a sudden modification of the harmonic potential, and we solve the corresponding time dependent equation, which takes the form of a generalized time-dependent Gross-Pitaevskii equation. We then monitor the evolution of the second moment $\langle x^2 \rangle$ of the in-trap density distribution, which, as shown in Fig.1b, presents a perfect oscillation. In the absence of the dipolar interaction, the mean-field solution for the breathing mode frequency can be analytically obtained by solving the hydrodynamic equations of superfluids[23] yielding $\omega = \sqrt{5/2}\,\omega_x$, in excellent agreement with experimental observations[25]. In the presence of dipolar interaction, we find a breathing frequency in agreement with hydrodynamic calculations[32], value which we show to extend beyond the mean-field region of stability of the BEC. The same value is obtained by a sum rule calculation, which provides a rigorous upper bound for the mode frequency (see Methods), as well as the solutions of the Bogoliubov equations[33,34].

For values of $\epsilon_{dd} \geq 1.38$, the equilibrium profiles exhibit a lattice structure, a signal of supersolidity[10], see Fig.1b. Remarkably, the corresponding time dependent solutions reveal a beating reflecting the presence of two oscillation modes. The nature of the two modes is identified through an analysis of the time evolution of the lattice period, which is dominated by the higher frequency mode, as well as of the lattice amplitude, dominated by the lower frequency mode (see Methods). We note that the bimodal oscillation emerges in a clean way only if we consider small amplitude oscillations (<5%) and tends to disappear for larger amplitudes, confirming that the two modes are strongly coupled, as easily seen from the hydrodynamic-like equations for the homogeneous gas[13]. When $\epsilon_{dd}$ is further increased into the droplet regime, the lower mode disappears and only the upper one is left.

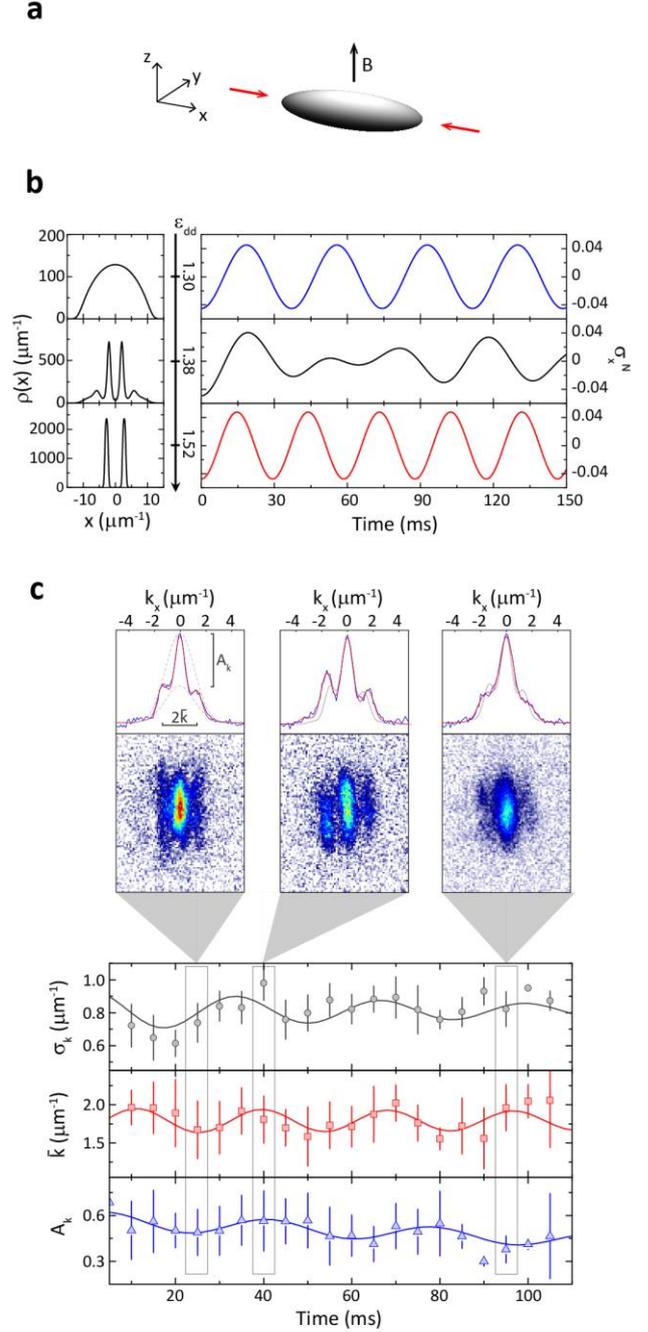

**Fig.1. Doubling of the axial breathing mode in a dipolar supersolid.** a) Geometry of the system: the axial breathing of the BEC leads mainly to a variation of the width along the x direction (arrows). b) Theoretical calculations of the stationary in-trap density distributions along the $x$ direction (left) and of the time-dependent oscillations in its normalized width, $\sigma_x^N(t) = \sqrt{\langle x(t)^2 \rangle / \langle x(0)^2 \rangle}$. Three representative cases are shown: standard BEC (top row, $\epsilon_{dd} = 1.30$), supersolid regime (middle row, $\epsilon_{dd} = 1.38$), and droplet crystal regime (bottom row, $\epsilon_{dd} = 1.52$). While BEC and crystal feature only one oscillation frequency, the supersolid regime shows clearly the beating of two frequencies, proving that the periodic modulation of the density is accompanied by the appearance of a new compressional mode. c) Experimental observation of the two oscillation modes in the supersolid regime ($\epsilon_{dd} = 1.38$), by monitoring the width of the distributions in momentum space, $\sigma_k(t) = \sqrt{\langle k_x(t)^2\rangle}$, (top panel, gray), the spacing of the side peaks $\bar{k}(t)$ (middle panel, red) and the modulation amplitude $A_k(t)$ (bottom panel, blue). Experimental data (circles, squares and triangles) are fitted with damped sinusoids (lines), to measure the dominant frequencies, which are $\omega/\omega_x$= 1.45(8), 1.66(10) and 1.27(12) for $\sigma_k$, $\bar{k}$ and $A_k$, respectively. The insets show samples of the experimental distributions in the $(k_x, k_y)$ plane and the related fits with a two-slit model to measure $\bar{k}$ and $A_k$; grey lines show the fit to the first distribution, for comparison. Error bars represent the standard deviation of 4-8 measurements.

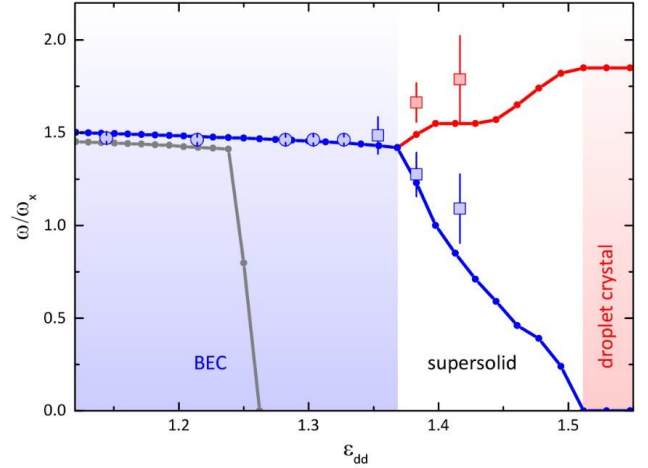

**Fig. 2. Axial mode frequencies from BEC to supersolid and droplet crystal.** Frequencies are normalized to the harmonic trap frequency in the $x$ direction; the dipolar character of the system is varied through the parameter $\epsilon_{dd} = a_{dd}/a_s$. Dotted lines are the theoretically predicted frequencies including the LHY energy term due to quantum fluctuations (blue and red) and excluding the LHY term (grey). Large circles and squares are experimentally measured frequencies: in the BEC regime (circles) the oscillation is induced by quenching $\epsilon_{dd}$; in the supersolid regime (squares) the roton instability naturally triggers the oscillation. Colors indicate the dominant character of the two modes in the supersolid: superfluid- (blue) or lattice-related (red). Error bars are one standard deviation. In the experiment, $\epsilon_{dd}$ has a calibration uncertainty of 3%.

In the experiment, differently from the theory, the system is prepared in the BEC regime and then slowly brought to the supersolid regime by increasing $\epsilon_{dd}$. We reveal the oscillations by monitoring the momentum distribution $n(k_x)$ along $x$, after a free expansion (see Methods). Simulating the initial expansion is challenging, so we cannot exactly relate the experimental and theoretical observables; however, the oscillation frequencies are not affected by the expansion. In the BEC regime, we can excite the axial breathing oscillation with a controlled amplitude by changing suddenly $a_s$. We monitor the second moment of the distribution, $\langle k_x^2\rangle$, whose oscillation features a weak damping due to finite-temperature effects. The same finite-temperature analysis confirms that the oscillation frequency of the BEC is distinct from that of a classical gas ($\omega = 2\omega_x$) and is therefore a consequence of the breaking of phase invariance (see Methods).

The transition to the supersolid regime is signaled by the appearance of side peaks in $n(k_x)$, which reflect the in-trap lattice[10], see Fig.1c. We observe that the roton instability itself triggers naturally the axial oscillation with a typical amplitude of 10%, which therefore represents the minimum amplitude in that regime. Due to the decay of the density modulation via three-body losses, the observation time is limited to few periods, reducing the precision of the measurements. Differently from the theory, in the time evolution of $\langle k_x^2\rangle$ we observe a single mode with a lower frequency than the BEC one. We can however monitor also two observables directly related to the two modes found in the theory: the spacing $\bar{k}$ of the side peaks, which is the inverse lattice period; the modulation amplitude $A_k$, which is associated to the depth of the lattice. As shown in Fig. 1c, the two observables feature clear sinusoidal oscillations, with different frequencies: $\bar{k}$ oscillates with a higher frequency and $A_k$ with a lower one, in agreement with the theoretical predictions.

In Figure 2 both the predicted and measured frequencies versus $\epsilon_{dd}$ are reported. The bifurcation of the BEC mode in the theory marks the onset of the supersolid regime. The spontaneous excitation of the axial mode observed in the experiment reveals a release of excess energy, suggesting that the transition is of the first order, as expected[1]. In the supersolid, the higher frequency mode is clearly related to the lattice deformations and its frequency increases due to the dipolar repulsion between neighboring density maxima. The lower mode is instead related to the compressional oscillation of the superfluid component; its downward frequency shift can be justified as an effective mass acquired by the atoms moving through the lattice, corresponding to a reduction of the superfluid fraction. The lower mode eventually disappears as the system enters in the droplet crystal regime, in analogy with the behavior of the corresponding Goldstone mode in uniform systems[1,13,16,21]. The upper mode instead approaches the value characteristic of a solid phase of incoherent droplets (see Methods).

The experiment-theory agreement on the BEC side is remarkable. The comparison with theoretical results without LHY term shows a clear stabilizing effect of quantum fluctuations already on the BEC side. In the supersolid regime, the splitting of the experimental oscillation frequencies agrees qualitatively with the theory. For larger $\epsilon_{dd}$, even before the theoretical prediction for the transition to the droplet crystal, the $\bar{k}$ and $A_k$ modes are no longer visible in the experiment because $n(k)$ becomes incoherent (see Methods).

In conclusion, the bifurcation of the lowest compressional mode of a harmonically-trapped dipolar supersolid gives evidence of the simultaneous breaking of two continuous symmetries, in analogy with the gapless Goldstone modes predicted for a homogeneous supersolid. The nature of the mode associated to the broken translational symmetry demonstrates the compressibility of the crystal structure of the dipolar supersolid, in analogy to the hypothesized He supersolid[1] and in contrast with the incompressible cavity supersolid[9,19,35]. The availability of the compressional modes opens also appealing new paths to explore the non-trivial competition between superfluid and crystalline features exhibited by this novel type of supersolids, in which two separate quantum phase transitions can be accessed by tuning a single interaction parameter. For example, a detailed analysis of the mode splitting at the BEC-supersolid transition might confirm that the transition is of the first order, in analogy to the discontinuity in the phonon velocities of the Goldstone modes[21]. Working with systems with reduced losses[12] and larger sizes, one might also explore the transition between the supersolid and the droplet crystal, where the theory predicts the disappearance of the lower frequency mode.

## METHODS

**Real time simulations.** Our theoretical predictions are based on the numerical solutions of a time dependent density functional equation, which takes the form of an extended Gross-Pitaevskii equation (eGPE) for the condensate wave function $\phi(r,t)$:

$$i\hbar \frac{\partial \phi(r,t)}{\partial t} = \left[ \frac{-\hbar^2}{2m}\nabla^2 + V_t(r) \right.$$
$$+ \int dr' (g\delta(r-r')$$
$$+ V_{dd}(r-r'))|\phi(r',t)|^2$$
$$\left. + \gamma(\epsilon_{dd})|\phi(r,t)|^3 \right] \phi(r,t)$$

Here, $V_t = \frac{1}{2}m(\omega_x^2 x^2 + \omega_y^2 y^2 + \omega_z^2 z^2)$ is the harmonic trapping potential and m the atomic mass. The contact interaction is expressed by: $g\delta(r-r')$, with $g = \frac{4\pi\hbar^2 a_s}{m}$, and $a_s$ the s-wave scattering length. $V_{dd}(r) = \frac{\mu_0 \mu^2}{4\pi} \frac{1-3\cos^2\theta}{|r|^3}$ is instead the dipole-dipole interaction between two magnetic dipoles aligned along the z axis, with $\mu$ the atomic dipole moment, $\mu_0$ the vacuum permeability and $\theta$ the angle between the vector $r$ and $z$. The dipolar parameter $\epsilon_{dd} = \frac{a_{dd}}{a_s}$ (where $a_{dd} = \frac{\mu_0 \mu^2 m}{12\pi\hbar^2}$ is the dipolar length) gives the relative strength of the (partially) attractive dipole-dipole interaction, and the repulsive contact interaction. The last term, $\gamma(\epsilon_{dd}) = \frac{32}{3\sqrt{\pi}} g a_s^{3/2} F(\epsilon_{dd})$, is the beyond-mean-field LHY correction, with $F(\epsilon_{dd}) = \frac{1}{2}\int d\theta \sin\theta [1 + \epsilon_{dd}(3\cos^3\theta - 1)]^{5/2}$ [29]. In the present density functional approach, the local atom density is identified as $n(r,t) = |\phi(r,t)|^2$.

The equilibrium density configuration is found by fixing the equation parameters to the values available in the experiments, and then evolving the eGPE in imaginary time. We consider N=35000 atoms of $^{162}$Dy, for which $a_{dd} = 130\, a_0$ ($a_0$ is the Bohr radius). The trapping frequencies are $\omega_{x,y,z} = 2\pi(18.5, 53, 81)$ Hz, and $a_s$ is changed to tune the value of $\epsilon_{dd}$.

The axial breathing mode is excited by finding the equilibrium configuration in the presence of a perturbation of the form $\hat{H} = -\lambda \hat{x}^2$, where $\lambda$ is a small parameter, and then evolving the eGPE in real time with $\lambda = 0$. During the evolution, we keep track of the time dependence of the second moment of the in-situ density distribution, defined as $\sigma_x(t) = \sqrt{\langle x^2(t)\rangle / \langle x^2(0)\rangle}$, with $\langle x^2(t)\rangle = \int dr\, x^2 n(r,t)$. This quantity shows a simple harmonic oscillation in the BEC and in the droplet crystal phase, while a beating between two frequencies clearly emerges in the supersolid regime.

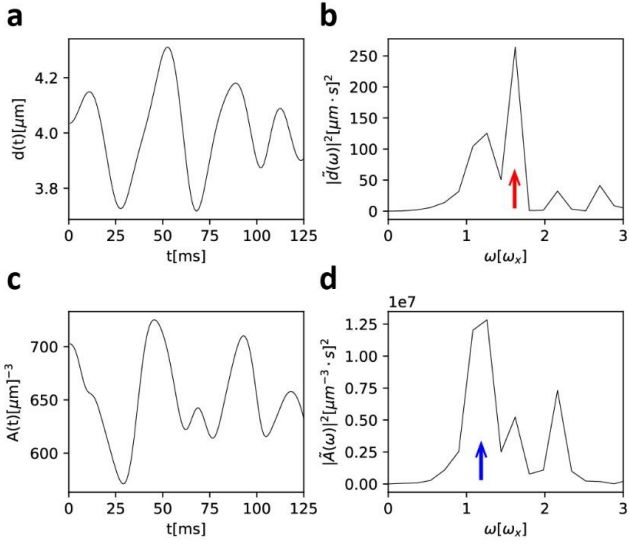

**Extended Data Fig.1. Peaks amplitude and relative distance.** In the supersolid phase, the axial breathing mode bifurcates into a higher and a lower frequency mode, mainly coupled respectively to the relative distance between the density peaks and to their amplitude. a) Time evolution of the relative distance $d(t)$ between two density peaks and b) its Fourier transform, dominated by a peak at the higher frequency (red arrow). c) Time evolution of the peak density amplitude $A(t)$ and d) its Fourier transform, dominated by the lower frequency mode (blue arrow).

**Theoretical mode assignment in the supersolid phase.** In the supersolid phase, we also record the time evolution of the relative distance $d(t)$ between the peaks of the in-situ density distribution, as well as their amplitude $A(t)$. The results are shown in Extended Data Fig.1 together with their Fourier transform $\tilde{d}(\omega)$ and $\tilde{A}(\omega)$, respectively. The dominant peak in $\tilde{d}(\omega)$ corresponds to the higher of the two frequencies that contribute to the beating observed in $\sigma_x(t)$, while the dominant peak in $\tilde{A}(\omega)$ correspond to the lower one. This allows us to identify the higher frequency mode as the one associated to a lattice deformation, and the lower frequency mode as the one associated with the compression of the superfluid.

**Sum rules.** A further insight on the excitation spectrum of the system is provided by the calculation of the sum rule $m_1/m_{-1}$ [26], where $m_1$ and $m_{-1}$ are, respectively, the energy weighted and inverse-energy weighted moments of the dynamic structure factor, relative to the operator $\hat{x}^2$. Explicit calculation of the two sum rules[36] gives the rigorous upper bound $\omega^2 = -2\frac{\langle x^2 \rangle}{d\langle x^2\rangle/d\omega_x^2}$ for the square of the frequency of the excited mode, which can be extracted from static calculations. In the BEC regime, we find that the upper bound matches with high precision the frequency of the axial breathing mode calculated by solving the Bogoliubov-de Gennes equations and by real time simulations, so that the linear response of the system to the operator $\hat{x}^2$ is exhausted only by this mode. In the supersolid regime, instead, the sum rule result lies between the two beating frequencies, so that in this phase the operator excites also the higher energy modes.

**Classical model for droplet oscillation.** For the value $\epsilon_{dd} = 1.53$, the equilibrium configuration of the system is given by two self-bound droplets, whose relative distance is fixed by the external trap. We find that the axial mode excited by $\hat{x}^2$ is just an out-of-phase harmonic oscillation of the two droplets without any significant deformation of their density profiles. To get a better physical insight on the nature of the oscillations in this regime, we study their frequency by treating the two droplets as classical distributions of dipoles. Neglecting their internal structure, we write the energy functional in the form $E[n_1, n_2] = E_{dd}[n_1, n_2] + E_{trap}[n_1] + E_{trap}[n_2]$, where $n_1(r)$ and $n_2(r)$ are the density distributions of the two droplets calculated by solving the eGPE, while $E_{dd}[n_1, n_2] = \frac{1}{2}\int d\,r_1 dr_2 V_{dd}(r_1 - r_2)n_1(r_1)n_2(r_2)$ is their dipole-dipole interaction energy and $E_{trap}n_i = \int d\,rV_t(r)n_i(r)$ is the energy term due to the trapping potential. By considering the variation of the energy functional for a small displacement of the center of mass of the droplets in the axial direction, one finds that the frequency of this out-of-phase oscillation is given by $\omega_x\sqrt{1 - \frac{2}{m\omega_{xN}^2}\int d\,r_1 dr_2 V_{dd}(r_1 - r_2)\frac{\partial n_1(r_1)}{\partial x}\frac{\partial n_2(r_2)}{\partial x}}$ [37]. This integral can be easily evaluated numerically and, for the choice $\epsilon_{dd} = 1.53$, gives the result $\omega/\omega_x = 1.95$, in relatively good agreement with the value $\omega/\omega_x = 1.85$ obtained from real time simulations.

**Experimental methods.** The experiments begin with a BEC of $^{162}$Dy atoms, with typical atom number $N=3.5\times10^4$, with no detectable thermal fraction, in a crossed beam dipole trap (see Ref. [38] for details). Typical trap frequencies are $\omega_x, \omega_y, \omega_z = 2\pi \times (18.5, 53, 81)$ Hz. Since there are day-to-day variations of the order of 5%, the trap frequencies are measured after each oscillation experiment.

The contact scattering length $a_s$ is controlled using a set of Feshbach resonances; we employ the precise $a_s(B)$ conversion provided in Ref. [39]. We calibrate the magnetic field amplitude through radio-frequency spectroscopy between two hyperfine states, with uncertainty ~1 mG [10]. The overall systematic uncertainty on the absolute value of $a_s$ is about 3 $a_0$, which corresponds to an uncertainty on $\varepsilon_{dd}$ of about 4%. To circumvent it, we identify a precise $B$ to $a_s$ conversion by comparing experimental and numerical data for the critical $\varepsilon_{dd}$ for the roton instability. The experimental resolution of $a_s$ is about 3 $a_0$. We checked, both experimentally and numerically, that the day-to-day variations of $a_s(B)$, of the trap frequencies and of the atom number produce a shift of the critical $\varepsilon_{dd}$ smaller than our resolution.

The condensate is initially created at $a_s = 140\ a_0$. The scattering length is then tuned with an 80 ms ramp to $a_s = 114\ a_0$, close to the roton instability, which occurs at $a_s \approx 94\ a_0$ ($\varepsilon_{dd} \approx 1.38$). Depending on the regime studied, we use different methods for exciting the axial breathing mode: 1) In the supersolid regime, the axial oscillation is naturally triggered by crossing slowly the phase transition from the BEC side, with a 30-ms long ramp in $a_s$. The typical oscillation amplitude of about 10% sets the minimum amplitude achievable experimentally in the supersolid regime. 2) In the BEC regime, we can excite a small-amplitude oscillation ($\approx$5%) by changing the scattering length from $a_s = 114\ a_0$ to the final value in 10 ms, see the point at $\varepsilon_{dd} = 1.35$ in Fig.2 and Extended Data Fig.2c. 3) For exciting large-amplitude oscillations in the BEC regime, we cross a narrow resonance located at approximately 5.5 G, see points below $\varepsilon_{dd} = 1.35$ in Fig.2 and Extended Data Fig.2e. By changing the duration of the crossing ramp, we can tune the amplitude of the oscillation in the range 20-50%.

After variable waiting time at the final scattering length, we record the atomic distribution in the *x-y* plane by absorption imaging after 95 ms of free expansion. Within 200 µs before the release of the atoms from the trapping potential, we set $a_s = 140\ a_0$, thus minimizing the effects of the dipolar interaction on the expansion[11]. We interpret the recorded distributions as momentum-space densities, $n(k_x, k_y)$. The imaging resolution is 0.2 µm$^{-1}$ (1/e Gaussian width).

**Transverse directions and higher modes.** The axial breathing mode is the lowest compressional mode of the dipolar BEC[32]. We have checked in both theory and experiment that it couples mainly to the width along the *x* direction, while the oscillations of the *y* and *z* widths have negligible amplitudes. The second compressional mode appears in the experiment only for large-amplitude oscillations of the lowest mode, it has a higher frequency and it couples mainly to the *y* width. For the small-amplitude oscillations studied experimentally in the supersolid regime, a very small oscillation of the *y* width is barely discernible, with a frequency seemingly close to the one along *x*. For the excitation procedure chosen in the theory, in the supersolid regime also higher compressional modes with larger frequencies can get excited (see Extended Data Fig.1). These modes couple mainly to the *y* and *z* widths and are therefore not relevant for the present analysis.

**Experimental fit procedures.** For each magnetic field and evolution time, we recorded between 10 and 20 distributions. For each measurement, we determine the 1D distribution $n(k_x)$ by integration along the *y* direction. We evaluate the second moment of $n(k_x)$ as $\sigma_k^2 = \sum_x (k_x - k_0)^2 n(k_x)$. In the supersolid regime, we use a double-slit fitting model[10]:

$$n(k_x) = C_0 e^{-(k_x-k_0)^2/2\sigma^2}\left[1 + C_1 cos^2\left((k_x - k_0)\pi/\bar{k} + \phi\right)\right].$$

The model describes two component: a gaussian distribution with a periodic modulation of period $\bar{k}$ and amplitude $C_0 C_1$, which is essentially the Fourier transform of the in-trap lattice; an unmodulated gaussian distribution of amplitude $C_0$, which resembles the measured distribution in $k$ space of an ordinary BEC and is related to the unmodulated component of the in-trap density. Given the complex expansion dynamics of dipolar quantum gases[40], it is not possible to relate analytically the distributions in $k$ space with those in real space. We however observe that the measured mean value of $\bar{k}$ agrees with the spatial frequency of the stationary supersolid density found in the theory. The modulation amplitude is defined as $A_k = C_1/(1 + C_1)$ and describes the weigth of the modulated component. In the analysis of the supersolid regime, we performed a selection excluding the data that does not show a clear stripe pattern: amplitude $C_1 < 0.35$ and fit error on $\bar{k} > 10\%$. The selected dataset was then used to provide data for $\sigma_k(t), \bar{k}$, and $A_k$ such as those shown in Fig.1.

**Experimental oscillations analysis.** For each magnetic field reported in Fig.2, we analyze the oscillation dynamics of $\sigma_k$ and, for the supersolid, of $\bar{k}$ and $A_k$. Each observable is fitted with a damped sinusoid,

$$X(t) = \Delta X sin\left(t\sqrt{\omega^2 - \tau^{-2}} + \varphi\right)e^{-t/\tau} + X_0 + \alpha t,$$

where $X(t)$ represents $\sigma_k(t)$, $\bar{k}(t)$ and $A_k(t)$, and $\Delta X, X_0, \omega, \tau, \varphi, \alpha$ are free fitting parameters. In particular, $\Delta X/X_0$ is the relative amplitude of the oscillation, $\omega$ the frequency and $\tau$ the damping time. A finite $\alpha$ accounts for possible slow drifts of the observables. For each dataset, we measured the trap frequency $\omega_x$ by exciting a collective dipole oscillation.

A clear oscillation of $\sigma_k(t), \bar{k}(t)$ and $A_k(t)$ is visible only for small-amplitude oscillations and in a narrow range of scattering lengths, $1.38 < \varepsilon_{dd} < 1.45$, corresponding to the supersolid regime. As shown in Extended Data Fig.2a, for $\varepsilon_{dd} = 1.50(5)$, close to the droplet regime, it is not possible to define $\bar{k}(t)$ and $A_k(t)$, because three-body recombination profoundly affects the dynamics of the system: dense droplets form and rapidly evaporate, destroying the lattice periodicity[10]. This corresponds to a fast loss in the atom number, which we measure experimentally. Extended Data Fig.2c-d shows two examples of small-amplitude oscillation of $\sigma_k(t)$. In the BEC regime, $\varepsilon_{dd} = 1.35(3), \sigma_k(t)$ oscillates as expected for a hydrodynamic superfluid: the value of the frequency $\omega/\omega_x = 1.48(10)$ is consistent with theory, see Fig. 2. Close to the droplet regime, $\varepsilon_{dd} = 1.50(5)$, $\sigma_k(t)$ oscillates with higher frequency, $\omega/\omega_x = 1.71(14)$.

In the BEC regime, we can precisely measure the mode frequency by exciting large amplitude oscillations, which allow to observe several periods. Extended Data Fig.2e-f shows two examples of large-amplitude oscillations. In the BEC regime, $\varepsilon_{dd} = 1.21(2)$, the oscillation frequency is $\omega/\omega_x = 1.46(2)$, in agreement with the theory. In the supersolid regime, $\varepsilon_{dd} = 1.38(2)$, large amplitude oscillations do not allow to distinguish clear oscillation of $\bar{k}(t)$ and $A_k(t)$. However, as shown in Extended Data Fig.2f, we observe a frequency shift in the oscillation of $\sigma_k(t)$ from short times to long times consistent with a decay of the system from the supersolid regime to the BEC regime in the first 150 ms, due to atom losses.

**Finite temperatures measurements.** Since finite temperatures can affect the axial breathing mode, we have performed a series of measurements in the BEC regime ($\varepsilon_{dd} = 1.14(2)$), where we can define an equilibrium temperature $T$. To obtain large temperatures, we interrupt the evaporative cooling procedure. We observe a small negative shift of the oscillation frequency for increasing $T$. Close to the critical BEC temperature, the shift from the measurements at the lowest temperature is $\Delta\omega/\omega \approx -10\%$. We observe also an increase of the damping rate for increasing $T$. Both phenomena can be justified as effects of the interaction of the condensate with the thermal component[41]. For a thermal gas above the condensation temperature we also observe the axial breathing mode, however with a completely different frequency, $\omega/\omega_x = 2.02(3)$, which is compatible with that expected for the breathing oscillation of a weakly-interacting classical gas, $\omega/\omega_x = 2$. These measurements demonstrate that our system is in the collisionless regime, so that the peculiar frequency observed for the axial breathing mode of the condensed component must be attributed to the locking of the wavefunction's phase[23]. Our knowledge of the lowest temperature achieved in the system is limited, since the temperature cannot be measured any longer when the thermal fraction drops below approximately 30%. Since the damping time decreases with $T$, we took care to maximise the damping time to 150-200 ms, in both BEC and supersolid regimes.

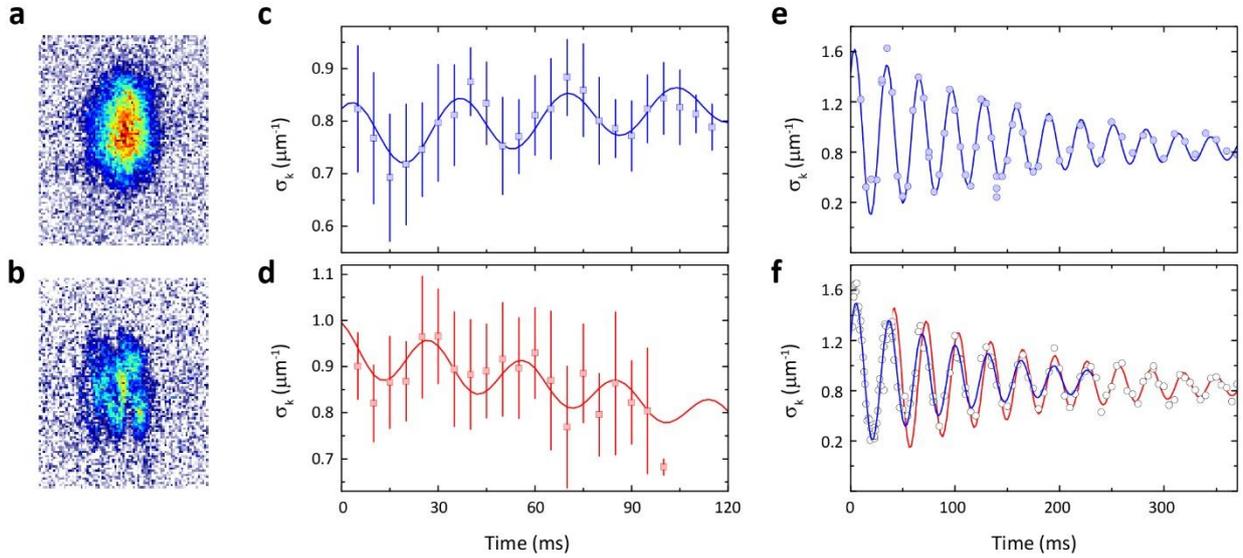

**Extended Data Fig. 2. Examples of small and large amplitude oscillations.** a-b): Typical experimental distributions in the $(k_x, k_y)$ plane for (a) the BEC regime close to the roton instability, $\varepsilon_{dd} = 1.35(3)$, and (b) the droplet regime, $\varepsilon_{dd} = 1.50(5)$. c-d): Small-amplitude oscillation of $\sigma_k(t)$ in the same conditions, (c) $\varepsilon_{dd} = 1.35(3)$ and (d) $\varepsilon_{dd} = 1.50(5)$. e-f): Large amplitude oscillation of $\sigma_k(t)$ for (e) the BEC regime, $\varepsilon_{dd} = 1.21(2)$, and (f) the supersolid regime, $\varepsilon_{dd} = 1.38(2)$. In the supersolid regime we observe a small frequency shift in the oscillation of $\sigma_k(t)$ during time: we fit $\omega/\omega_x = 1.43(1)$ up to about 150 ms (blue line), and $\omega/\omega_x = 1.48(2)$ from 150 ms on (red line).

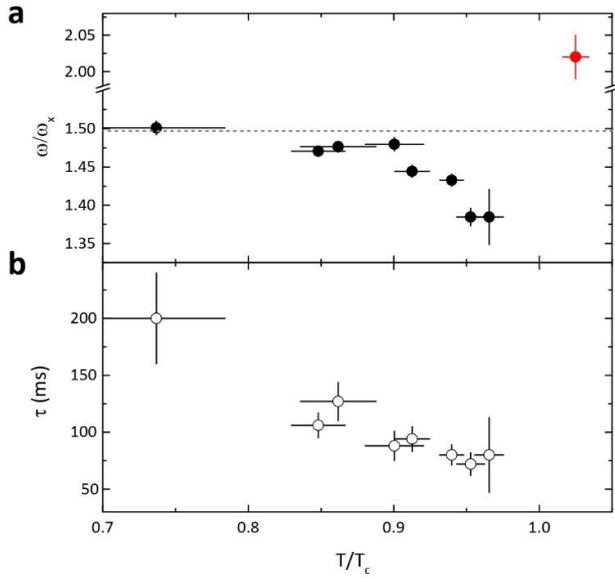

**Extended Data Fig.3. Finite temperature measurements.** a) Measured frequency and b) measured damping time of the axial breathing mode in the BEC regime for increasing temperature. The measured frequency for a thermal gas (red dot) is compatible with $\omega/\omega_x = 2$, demonstrating the collisionless nature of the system and the breaking of phase invariance for the BEC.

**Acknowledgements.** This work received funding by the EC-H2020 research and innovation program (Grant 641122 - QUIC). We acknowledge discussions with J.G. Maloberti and technical assistance by A. Barbini, F. Pardini, M. Tagliaferri and M. Voliani. S.M.R., A.R. and S.S. acknowledge funding from Provincia Autonoma di Trento and the Q@TN initiative. We acknowledge useful discussions with the participants to the Stuttgart meeting on "Perspectives for supersolidity in dipolar droplet arrays", where we became aware of related work by the groups of T. Pfau and F. Ferlaino.

**Author contributions.** L.T., E.L., F.F., A.F., C.G. and G.M. conceived the experimental investigation, performed the measurements and the experimental data analysis. S.M.R., A.R. and S.S. conceived the theoretical investigation, performed the simulations and the theoretical data analysis. All authors contributed to discussions and writing of the paper.

**Competing interests.** The authors declare no competing interests.

**Materials & Correspondence.** Correspondence and requests for materials should be addressed to modugno@lens.unifi.it and alessio.recati@unitn.it.